\documentstyle[epsfig]{aipproc}

\begin{document}
\title{Extracting Neutron Star Properties from X-ray Burst Oscillations}

\author{Nevin Weinberg$^*$, M. Coleman Miller$^{\dagger}$, and
Donald Q. Lamb$^*$}
\address{$^*$University of Chicago, Department of Astronomy and 
Astrophysics,\\ 
5640 South Ellis Ave., Chicago, IL 60637\\
$^{\dagger}$University of Maryland, Department of Astronomy, 
College Park, MD 20742-2421}

\maketitle

\begin{abstract}
Many thermonuclear X-ray bursts exhibit brightness oscillations.  The
brightness oscillations are thought to be due to the combined effects
of non-uniform nuclear burning and rotation of the neutron star.  The
waveforms of the oscillations contain information about the size and
number of burning regions.  They also contain substantial information
about the mass and radius of the star, and hence about strong gravity
and the equation of state of matter at supranuclear densities.  We have
written general relativistic ray-tracing codes that compute the
waveforms and spectra of rotating hot spots as a function of photon
energy. Using these codes, we survey the effect on the oscillation
waveform and amplitude of parameters such as the compactness of the
star, the spot size, the surface rotation velocity,
and whether there are one or two spots.
We also fit phase lag versus photon energy curves to data from the
millisecond X-ray pulsar, SAX~J1808--3658.
\end{abstract}

\section*{Introduction}

Shortly after the launch of the {\it Rossi} X-ray Timing Explorer 
(RXTE) in late 1995, single kilohertz brightness oscillations were discovered  
in RXTE countrate time series data from thermonuclear X-ray bursts in
several neutron-star low-mass X-ray binaries.  These oscillations are
remarkably coherent and their frequencies are very stable from burst to
burst in a given source \cite{SSZ98}.  
These oscillations are therefore thought to be at
the stellar spin frequency or its first overtone.  This suggests that
the oscillations are caused by rotational modulation of a hot spot
produced by non-uniform nuclear burning and propagation.  Analysis of
these oscillations can therefore constrain the mass and radius of the
star and yield valuable information about the speed and type of thermonuclear
propagation.  In turn, this has implications for strong gravity and dense
matter, and for astrophysical thermonuclear propagation in other contexts,
such as classical novae and Type Ia supernovae.

A comparison of theoretical waveforms with the observations is required
to extract this fundamental information.  Here we exhibit waveform
calculations that we have produced using general relativistic ray-tracing
codes.  We survey the effects of parameters such as the spot size, the
stellar compactness, and the stellar rotational velocity,
and demonstrate that our computations
fit well the phase lag data from SAX~J1808--3658.

\section*{Computational Method}

To compute observed light curves, we do general relativistic ray tracing
from points on the surface to the observer at infinity in a way similar
to, but more general than, \cite{PFC83} and \cite{ML98}.  For simplicity,
we assume that the exterior spacetime is Schwarzschild, that the surface
is dark except for the hot spot or spots, and that there is no background
emission.  The amplitudes would be reduced by a constant factor
if there were background emission. The angular dependence of the specific
intensity at the surface depends on both radiation transfer effects and
Doppler boosting (see \cite{WML99}).
For each phase of rotation we compute the projected area of many small
elements of a given finite size spot. We then build up the light curve of
the entire spot by superposing the light curve of all the small elements.
The grid resolution of the spot is chosen so that the effect of having a
finite number of small elements can alter the value of the computed
oscillation amplitudes by a fraction no larger than $\sim 10^{-4}$.  
After computing the oscillation waveform
using the above approach, we Fourier-analyze the resulting light curve to
determine the oscillation amplitudes and phases as a function of photon
energy at different harmonics.

\section*{Results}

\begin{figure}[ht]
\epsfig{file=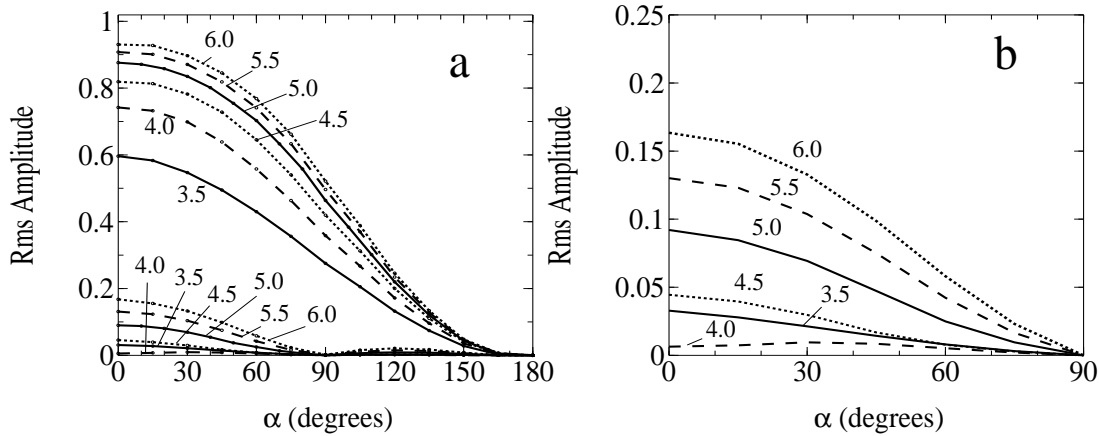,height=3.0truein, width=6.0truein}
\vskip 0.2truein
\caption[]{\label{fig1}
(a)~Bolometric rms amplitude vs.\ the angular radius $\alpha$ of the spot
at the first
harmonic (upper curves) and the second harmonic (lower curves)
from a single emitting spot centered on the rotational equator as
seen by a distant observer in the rotational
plane.  Numbers denote values of $R/M$, where we use geometrized
units in which $G=c\equiv 1$. (b)~Rms amplitude vs.
$\alpha$ at the second harmonic from two antipodal emitting
spots, with both the spots and the distant observer in the rotational
plane.  Note the change in vertical scale in panel (b).
In both cases we assume a nonrotating star and an isotropic specific 
intensity as measured by a local comoving observer.}
\end{figure}

\begin{figure}[ht]
\begin{minipage}[t]{2.8truein}
\mbox{}\\
\epsfig{file=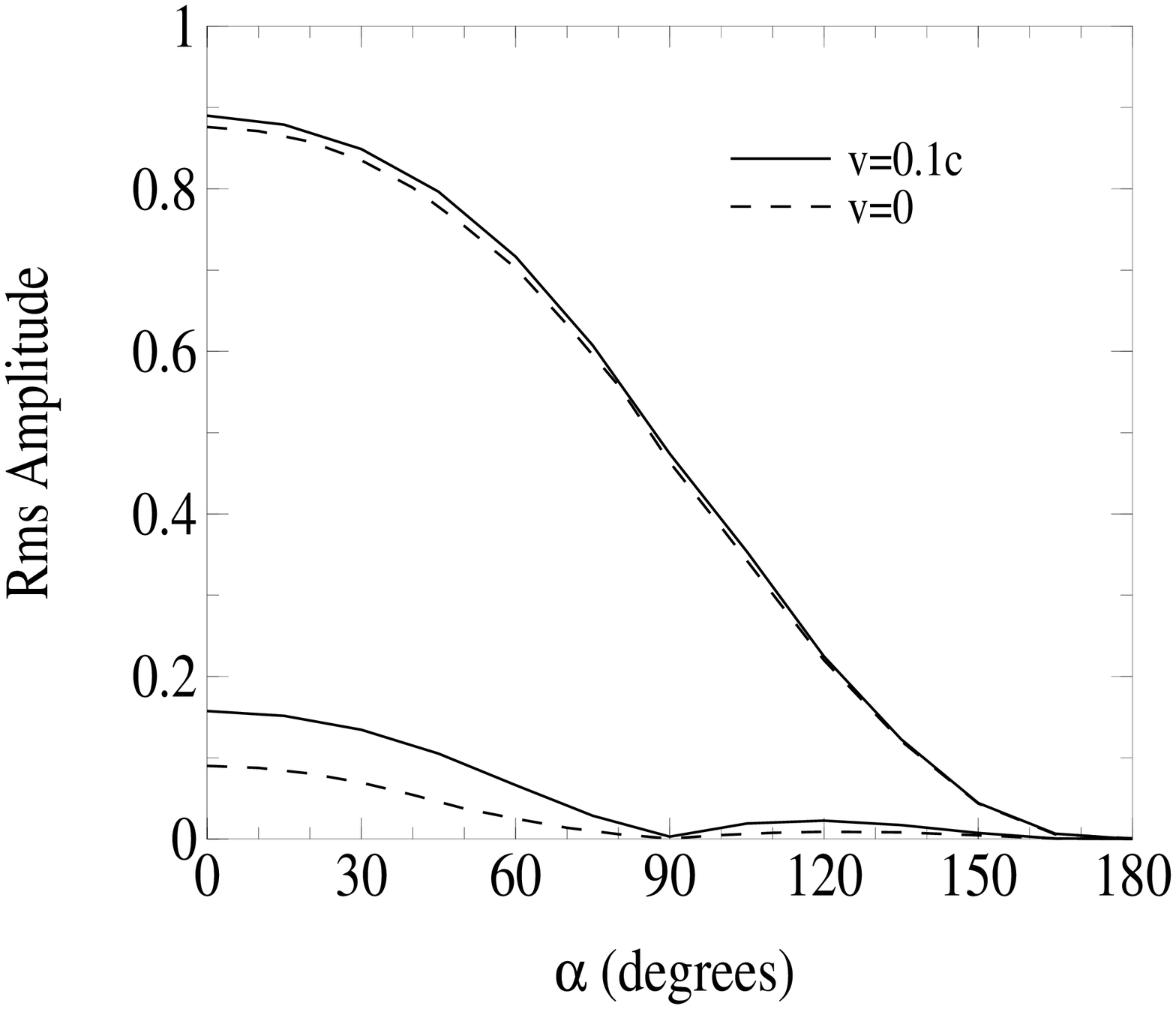, height=3.0truein, width=3.0truein}
\end{minipage}
\begin{minipage}[t]{2.8truein}
\mbox{}\\
\vskip 0.3truein
\epsfig{file=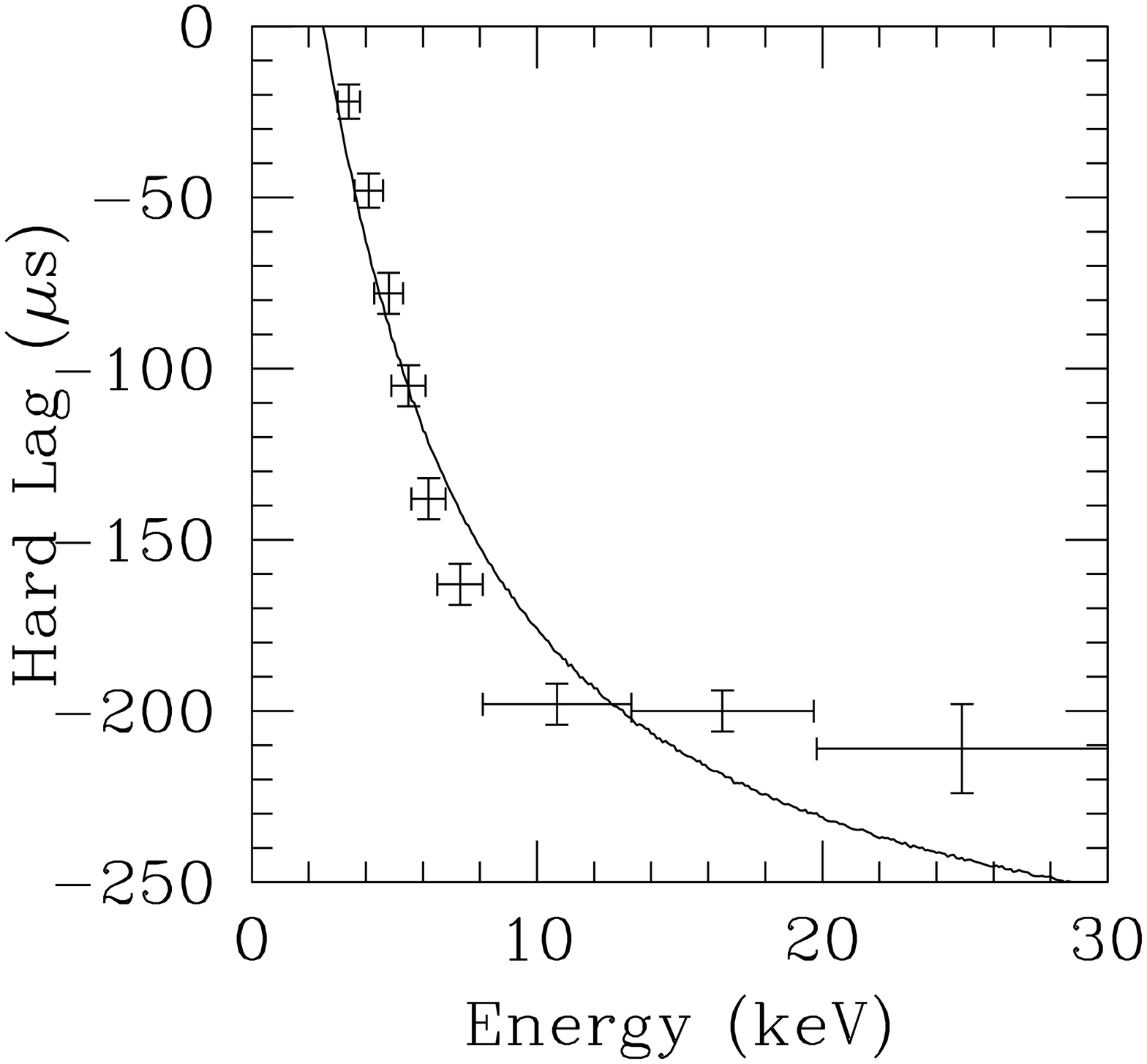, height=2.7truein, width=2.7truein}
\end{minipage}
\vskip 0.2truein
\caption[]{\label{fig23}
(left panel) Rms amplitude versus spot angular radius $\alpha$
at the first harmonic (upper curves) and the  second
harmonic (lower curves) from a single emitting spot with $v=0.1c$
(\textit{solid lines}) and $v=0$ (\textit{dashed lines}), $R/M = 5.0$, and
a spot and a distant observer both in the rotational equator. (right panel)
Phase lags versus photon energy for the millisecond X-ray
pulsar SAX~J1808--3658.  The data (crosses) are from \cite{CMT98},
where the reference energy is the 2--3~keV band.  The solid
line shows the phase lags in a Doppler shift model, assuming a
gravitational mass of $1.8\,M_\odot$ and a rotational velocity
of 0.1~c as measured at infinity, which would be appropriate for
the observed 401~Hz spin frequency and $R=11$~km.
The angular and spectral emission at the surface are that of a
grey atmosphere with an effective temperature of 0.6~keV as measured
at infinity.  The excellent
fit apparent in this figure supports the Doppler shift
explanation for the soft lags in this source.}
\end{figure}

Panel (a) of Figure~1 shows the fractional rms amplitudes at the
first two harmonics as a function of spot size and stellar compactness
for one emitting spot
centered on the rotational equator, as seen by a distant observer in
the rotational plane.  These curves demonstrate that a
common result of the hot-spot model is large-amplitude brightness oscillations
with a high contrast in strength between the dominant harmonic and weaker
harmonics, as is observed in several sources.
The curves for the first harmonic illustrate the
general shape of most of the first harmonic curves. Initially, the amplitude
depends only weakly on spot size. However, once the spot grows to an angular
radius of $\sim
40^{\circ}$ there is a steep decline in the oscillation amplitude which
flattens out only near the tail of the expansion.  This expected behavior
appears to be in conflict with the decline in amplitude observed by
Strohmayer, Zhang, \& Swank (1997) from 4U~1728--34, in which the initial 
decline is steep.  The cause of this could be that the initial velocity
of propagation is large, or that the observed amplitude is diminished
significantly by isotropization of the beam due to scattering (Weinberg,
Miller, \& Lamb 1999).  

Panel (b) of Figure~1 shows the
fractional rms amplitude at the second harmonic under the same assumptions but
for two identical, antipodal emitting spots. The range in spot size here is
$0^{\circ}-90^{\circ}$ since two antipodal spots of $90^{\circ}$ radii cover
the entire stellar surface. Note that in this situation, there is no first
harmonic.  

These figures show that when there is only one emitting spot,
the fundamental is always much stronger than higher harmonics.  Thus, a source 
such as 4U~1636--536 with a stronger first overtone than fundamental 
\cite{M99} must have
two nearly antipodal emitting spots.  As described in detail in \cite{WML99},
we confirm the results of \cite{PFC83} and \cite{ML98}
that the rms amplitude decreases  with increasing compactness
until $R/M\approx 4$, then increases due to the formation of caustics.
We also find that a finite surface rotational velocity increases the
amplitude at the second harmonic substantially, while having a relatively
small effect on the first harmonic (left panel of Figure~2).

As an application to data, in the right panel of Figure~2 we use our models
to fit phase lag data from the millisecond accreting X-ray pulsar 
SAX~J1808--3658.  The waveforms from the accreting spot are expected to
be similar to the waveforms from burst brightness oscillations, and the
signal to noise for this source greatly exceeds that from burst sources
such as Aql~X-1 \cite{F99}.  As is apparent from the
figure, the fit is excellent.  Further data, especially from a high-area
timing mission, could be used to constrain the stellar mass or radius
from phase lag data.

\end{document}